\documentclass[prb,aps,preprint]{revtex4-2}
\usepackage{graphicx}  
\usepackage{dcolumn}   
\usepackage{bm}        
\usepackage{amssymb}
\usepackage[english]{babel}
\usepackage{hyperref}
\usepackage{relsize}
\usepackage{titlesec}
\hypersetup{
	colorlinks=true,
	linkcolor=blue,
	filecolor=blue,      
	urlcolor=blue,
}
\begin{document}

\title{Synthesis and EOS study of orthorhombic (Fe,Ni)$_{7}$(C,Si)$_{3}$ and its importance as a possible constituent of Earth's core}

\author{Bishnupada Ghosh$^1$,  Mrinmay Sahu$^1$, Pinku Saha$^{1,2}$, Nico Giordano$^3$ and Goutam Dev Mukherjee$^1$ }

\address{$^1$ Department of Physical Sciences, Indian Institute of Science Education and Research Kolkata, Mohanpur Campus, Mohanpur 741246, Nadia, West Bengal, India.}
\address{$^2$ Department of Earth Sciences, ETH Zürich, Zurich 8092, Switzerland.}
\address{$^3$ Photon Science, Deutsches Elektronen Synchrotron, Hamburg, Germany.}
\begin{abstract}
 We have synthesized an orthorhombic phase of nickel and silicon doped Fe$_{7}$C$_{3}$ at high-pressure and high temperature using a laser-heated diamond anvil cell. The synthesized material is characterized using X-ray diffraction (XRD), Raman spectroscopy, and Transmission Electron Microscopy (TEM) measurements. High-pressure XRD measurement at room temperature up to around 121 GPa is performed. The anomaly observed in the pressure evolution of unit cell volume around 79 GPa along with a slight elastic softening might be associated with a magnetic transition present in the material. The estimated bulk modulus shows a higher value due to the presence of less compressible nickel in the material. Density at core condition is calculated from the thermal pressure corrected equation of state (EOS), which gives an excellent match with the PREM data.   
\end{abstract}

\date{\today}
\maketitle

\section{Introduction}
The composition of the Earth's solid inner core is a matter of controversy till now. Earth's solid inner core consists of  iron with some amount of nickel as reported by several cosmochemical, geochemical, and geophysical studies (see\cite{dubrovinsky2007body,lin2002iron,birch1952elasticity}, and references therein). But the higher densities of iron, nickel as well as Fe-Ni alloys at core pressure and temperature conditions compared to the values obtained from seismic observations, suggest the presence of light elements in the core \cite{dziewonski1981preliminary, stevenson1981models}. Carbon, silicon, sulfur, and oxygen are the proposed light elements in the  core \cite{cote2008light, allegre2001chemical}. Among the proposed light elements, carbon is a strong candidate as it is extremely abundant in the solar system and poses a strong affinity to solid as well as liquid iron \cite{wood1993carbon}.
Previous high-pressure studies show that some intermediate iron-carbon (Fe-C) compounds, like Fe$_{3}$C and Fe$_{7}$C$_{3}$ can be considered as the main carbon-bearing phases in the core\cite{liu2016phase}. Wood\cite{wood1993carbon} proposed that Fe$_{3}$C can be the primary component of Earth's inner core, which is found to be stable up to 187 GPa at 300 K and 73 GPa at 1500 K, respectively, from X-ray diffraction(XRD) studies\cite{scott2001stability,li2002compression,sata2010compression}. But according to Dasgupta et al.\cite{dasgupta2009high} and Nakajima et al.\cite{nakajima2009carbon}, Fe$_{3}$C melts to Fe-C alloy and more carbon-rich Fe$_{7}$C$_{3}$ while temperature quenching above 5–10 GPa in the multi anvil press experiments. Most of the recent experimental and theoretical studies on the Fe-C system indicate that Fe$_{7}$C$_{3}$ can become more stable than Fe$_{3}$C as a carbon-bearing phase in the core\cite{lord2009melting,mookherjee2011high,nakajima2011thermoelastic}.
At ambient conditions, Fe$_{7}$C$_{3}$ may have two crystal structures. First a hexagonal polytype (h-Fe$_{7}$C$_{3}$) with the space group P6$_{3}$mc\cite{mookherjee2011high}\cite{nakajima2011thermoelastic}\cite{chen2014hidden} and second an orthorhombic polytype (o-Fe$_{7}$C$_{3}$) with possible space groups Pnma, Pmcn, and Pbca\cite{eckstrom1950new,herbstein1964identification,fruchart1969twin,tsuzuki1984high,litasov2015thermal,prescher2015high,lai2018high}.

Besides, this iron-carbide system can occur in pure form as well as in phases containing a few percentages of light elements as impurities. As the overlying mantle is very rich in silicate minerals, silicon is considered to be one of the most possible impurities present in iron carbide. The density of pure Fe$_{7}$C$_{3}$ at core pressure and temperature conditions is found to be less than the PREM data\cite{dziewonski1981preliminary, liu2016phase}. Das et al. \cite{das2017first} have theoretically shown that doping a few weight percent of Si impurities at the carbon sites in Fe$_{7}$C$_{3}$ carbide phases, the values of Poisson’s ratio and density increase. They also found that the agreement with PREM data is better for the orthorhombic phase of iron carbide (o-Fe$_{7}$C$_{3}$) compared to the hexagonal phase (h- Fe$_{7}$C$_{3}$). But still the density of Si-doped Fe$_{7}$C$_{3}$ is less than the PREM data at core condition. In an experimental work Saha et al. \cite{saha2021synthesis} has synthesized silicon doped Fe$_{7}$C$_{3}$ using laser-heated diamond anvil cell (LHDAC) and found the crystal structure to be orthorhombic. Their density estimation gives a closer value to PREM data compared to the theoretical prediction by Das et al.\cite{das2017first}.
A recent theoretical study by Chatterjee et al.\cite{chatterjee2021ni} shows that the addition of 6.25\% Ni in Fe stabilizes the bcc phase of Fe-Ni alloy at core pressure and temperature. The density of Fe-Ni-Si alloy estimated by Morrison et al. at core pressure and at 5500K matches very well with the PREM data\cite{morrison2018equations}.
As one can not neglect the presence of carbon in the core composition, in the present study we have synthesized an orthorhombic phase of (Fe,Ni)$_{7}$(C,Si)$_{3}$, from a powder mixture of iron, nickel, carbon and silicon at high pressure and high temperature. We characterized the synthesized sample using XRD and later using transmission electron microscopy(TEM) and Raman measurements. A high-pressure XRD study of the synthesized material is carried out up to about 120 GPa at room temperature. The density of the material, calculated by extrapolating the thermal pressure corrected EOS at 7000 K up to core pressure (364 GPa)  is found to be close to the PREM data.  
\\
\section{Experimental methodology}

A powder mixture of iron(CAS No: 7439-89-6, purity: $>$ 99.5\% (RT)), nickel(CAS No:7440-02-0, purity:99.9\%), diamond nanopowder (CAS No:7782-40-3, purity $\geq$ 95\%) and silicon(CAS No: 7440-21-3, purity:99\%)(stoichiometric ratio Fe:Ni:C:Si = 6.56:0.44:2.50:0.50) is mixed well and ground in an alcohol medium for about 12 hours, to get a homogeneous mother material for the synthesis. The percentage of silicon doping at the C site (16.7\%) is considered from the experimental work of Saha et al. \cite{saha2021synthesis} and the nickel doping percentage at the Fe site (6.25 \%) is taken from the predictions of theoretical work by Chatterjee et al. \cite{das2017first}. 

The powder mixture of the sample is pressed using a 20-ton manual hydraulic press to make around 10 $\mu$m thin flake of the sample. To synthesize the sample, it is heated at a temperature of around 2400K and a pressure of around 43 GPa, using the laser-heated diamond anvil cell (LHDAC) facility in our laboratory as well as at the P02.2 beamline at the DESY synchrotron facility.
Symmetric type diamond anvil cell (DAC) with diamonds having culet diameter 100 $\mu$m is used to compress the sample loaded in a 50 $\mu$m hole drilled in a Re gasket preindented to a thickness of 25 $\mu$m. MgO is used as a pressure transmitting medium (PTM) as well as a pressure marker. Oven-dried MgO powder is pressed to make thin dense flakes. Then small pieces of these flakes are placed inside the gasket hole on the lower diamond anvil. Then a thin sample flake is loaded on the top of the MgO. More flakes of MgO are placed on the top of the sample. This MgO-sample-Mgo sandwich sample geometry helps to insulate the sample from the diamond anvil. The pressure inside the cell is increased gradually up to around 43 GPa using the gas membrane attachment. 
The high-energy 100-watt NdYag laser (1064 nm) is used to heat the sample from both sides of the DAC. The incandescent light emitted from the hotspot of the sample is collected and fitted to the Planck model to estimate the temperature. The loaded sample is at the center of the hole with a well-defined separation from the gasket edge (shown in the inset of Fig.3(a)). The whole sample is heated around 2400 $\pm$ 100 K by moving the laser hotspot position throughout the whole sample surface loaded inside the symmetric LHDAC. The XRD pattern at 43 GPa before heating of the sample is collected using the monochromatic X-ray beam of 0.2910 $\AA$ wavelength available in the beamline. The beam cross-section is 1.2$\mu$m$\times$2.3$\mu$m approximately. 2D X-ray diffraction patterns are collected using a Perkin-Emler 1681 detector aligned normal to the beam. The sample-to-detector distance is calibrated from the 2D XRD pattern of CeO$_{2}$.  Acquired 2D XRD images are converted into 2$\theta$ vs intensity plot using DIOPTAS software\cite{prescher2015dioptas}. Pressure is calibrated using 3$^{rd}$ order Birch Murnaghan equation of state (BM EOS) of MgO\cite{speziale2001quasi}. X-ray diffraction patterns are indexed using the CRYSFIRE software\cite{shirley1999crysfire} followed by Lebail fitting using the GSAS software\cite{toby2001expgui}. 
Also, the sample is similarly synthesized using an in-house laser-heated diamond anvil cell facility at IISER Kolkata. The pressed sample flakes are loaded in a plate-type Bohler-Almax DAC having diamonds of 300 $\mu$m culet diameter. A hole of diameter 100 $\mu$m is drilled on a steel gasket preindented to 30 $\mu$m. The sample is loaded inside the hole using MgO PTM and ruby pressure marker. The sample is first pressed to 43 GPa and then heated to about 2400$\pm$ 100K using the 100-Watt Ytterbium fiber laser by IPG Photonics (Model No. YLR-100-SM-AC-Y11). The temperature is calculated by fitting the collected incandescent spectrum to the Planck model using grey body approximation\cite{mukherjee2007high,saha2021temperature}. After quenching to room temperature the heated sample is retrieved. The recovered sample is placed on a Cu-grid and used for selected area electron diffraction (SAED) using HR TEM operating at 200 kV. 
The Raman spectroscopic measurements are also performed on the parent sample powder and the recovered quenched sample at ambient pressure and room temperature, for confirmation of the synthesis of the new phase.
 
\section{Results and Discussions}
\subsection{Synthesis and X-ray diffraction measurements}
We first discuss the results of the synthesis carried out in our lab. The Raman spectra of the parent mixture sample and the synthesized material are shown in Fig.1(a). The Raman spectrum of the parent sample mixture before heating shows the prominent silicon peak (around 520 cm$^{-1}$) as well as D-band (1334 cm$^{-1}$) and G-band (1550 cm$^{-1}$) of diamond nanopowder. But the absence of those Raman peaks in the spectrum of the heated sample indicates the absence of unreacted elements and the formation of a new phase. The SAED pattern of the recovered heated sample is shown in Fig.1(b). Indexing the SAED pattern gives an orthorhombic structure with $a$=12.7011(8) \AA , $b$=3.5712(5) \AA , $c$=14.9324(11) \AA , unit cell volume V= 677.30(3) \AA$^{3}$. Room temperature XRD images of the sample loaded in LHDAC before heating and after heating (quenched to the room temperature) are collected from the same part of the sample and are shown in Fig.2(a-b). The 2D XRD image of the quenched sample shows spotty nature and several new diffracted rings indicating the formation of small crystallites. The XRD pattern obtained from the image after heating is indexed to an orthorhombic crystal structure having space group \textit{Pbca} with lattice parameters $a$=11.3519(10) \AA , $b$=4.1824(3) \AA , $c$=13.2964(13) \AA , unit cell volume V= 631.29(10) \AA$^{3}$ and Z=8 (Fig.3(a)). The indexed lattice parameters are very close to the orthorhombic crystal structure of pure Fe$_{7}$C$_{3}$ at 46 GPa, reported by Preshcher et al. \cite{prescher2015high}  and Fe$_{7}$(C,Si)$_{3}$ at 49 GPa reported by Saha et al.\cite{saha2021synthesis}. The LeBail fit of the XRD pattern of the synthesized sample using the indexed orthorhombic structure gives an excellent fit and is shown in Fig.3(a). 
Since Fe$_{7}$C$_{3}$ is found to show two polymorphs, hexagonal\cite{mookherjee2011high,nakajima2011thermoelastic,chen2014hidden} and orthorhombic \cite{eckstrom1950new,herbstein1964identification,fruchart1969twin,tsuzuki1984high,litasov2015thermal,prescher2015high,lai2018high}, we have tried to match our XRD pattern to the hexagonal crystal structure. We have carried out the LeBail profile fit of the obtained XRD pattern using the hexagonal crystal structure of Fe$_{7}$C$_{3}$ as reported by Chen et al.\cite{chen2012magneto} as a starting model. However we fail to get a good fit of the XRD pattern as the major intense peak can not be fitted as shown in Fig.3(b). Therefore we proceed for further analysis of our data using the orthorhombic structure. The unit cell volume is found to be 1.8 \% less than tha unit cell volume at 43 GPa of Fe$_{7}$(C,Si)$_{3}$ synthesized by Saha et al. \cite{saha2021synthesis}. The decrease in unit cell dimension might be due to the 6.5 \% doping of nickel at iron site, which has a smaller atomic volume compared with Fe\cite{hirao2022equations}. 

The pressure evolution of the XRD patterns of the synthesized sample  up to about 121 GPa is shown in Fig.4. No major change in the XRD pattern is found in the pressure region studied, indicating the absence of any structural phase transition. 
The pressure evolution of relative lattice parameters is shown Fig.5(a).
From the figure, it is evident that $a$-axis is compressed most in comparison to other axes. Also $b$ and $c$-axes show plateau-type anomaly in the pressure range 70-80 GPa. $a$ axis is reduced by about 5.9 \% where $b$ and $c$ axes are compressed by about 3.7 \% and 4.2 \%, respectively up to 121 GPa. Prescher et al. \cite{prescher2015high} showed that o-Fe$_{7}$C$_{3}$ exhibit a paramagnetic to non-magnetic transition around 70 GPa. Though they didn't find any discontinuity or slope change in the pressure evolution of the volume data, the mean Fe-C distance shows slope changes around 70 GPa. The variation of unit cell volume with applied pressure is shown in Fig.5(b).
The pressure vs unit cell volume data shows a slight slope change around 78.9 GPa. The pressure vs volume data can not be fitted  using a single equation of state in the whole pressure region studied. Therefore, it is fitted using two separate 3rd order Birch Murnaghan equations of state (BM EOS) in two pressure regions (i) 43-78.9 GPa and (ii) 78.9-121 GPa, respectively. At first, the first region is fitted using 2$^{nd}$ order BM EOS by keeping K$^{'}$ fixed to 4.0. Which gives V$_{0}$= 705.14(3) \AA$^{3}$. Next the same region is fitted using 3$^{rd}$ order BM EOS by keeping V$_{0}$ fixed to 705.14(3) \AA$^{3}$ and varying both K$_{0}$ and K$^{'}$ which improved the chi square. The first fit gives K$_{0}$=303(14) GPa and K$^{'}$=4.1(6) .
To fit the second region using 3$^{rd}$ order BM EOS we have fixed the V$_{0}$ same as 1$^{st}$ region due to the absence of any structural phase transition within this pressure regime. This gives K$_{0}$=286(23) GPa and K$^{'}$=4.2(7). The slight decrease in the bulk modulus occurs at the pressure regime very close to the pressure point where paramagnetic to non-magnetic transition is observed by Prescher et al.\cite{prescher2015high}.  In a similar study by Saha et al. \cite{saha2021synthesis} pressure-induced elastic softening is found in Si-doped Fe$_{7}$C$_{3}$ at around 78 GPa. In previous studies on Fe$_{7}$C$_{3}$ elastic softening is related to spin transitions\cite{mookherjee2011high,nakajima2011thermoelastic}. According to Chen et al.\cite{chen2012magneto} the high-pressure discontinuity in the compression curve near 53 GPa is interpreted as a result of the high-spin to low-spin transition of Fe present in h-Fe$_{7}$C$_{3}$.
In our study, a slight drop in bulk modulus is found around 78.9 GPa, which might be a sign of a magnetic transition in the sample. 

The pressure evolution of bulk modulus found in our study is compared with other studies on pure and doped Fe$_{7}$C$_{3}$ and is shown in Fig.6(a). The studies by Saha et al.\cite{saha2021synthesis} and Chen et al.\cite{chen2012magneto} show sharp drops in bulk modulus at about 78 GPa and 53 GPa respectively.  The pressure evolution of K in our study matches quite well with the study by Prescher et al.\cite{prescher2015high}. The decrease in K(P) around 79 GPa in our study is about 2.2\%. This is smaller than those reported by Saha et al. on o-Fe$_{7}$(C,Si)$_{3}$ (42\% decrease). Replacing Fe by about 6.25 \% of Ni, probably leads to the elastic softening with a lesser degree, as Ni does not show any magnetic transition in this pressure range.
The pressure variation of the bulk modulus value of our study is compared with respect to PREM data by extrapolating up to core pressure (364 GPa) at room temperature (Fig.6(b)). Our study estimates a 27\% higher value of bulk modulus at room temperature with respect to the PREM data. Whereas extrapolated bulk modulus observed by Prescher et al.\cite{prescher2015high} gives a 35\% higher value than observed in PREM data. The bulk modulus of h-Fe$_{7}$C$_{3}$ estimated by Nakajima et al. \cite{nakajima2011thermoelastic,chen2012magneto} and Chen et al. are found to be 9.5\% higher and 15.5 \% less than the PREM data respectively, at core pressure. A similar study by Saha et al. on silicon doped Fe$_{7}$C$_{3}$ in its orthorhombic phase exhibits 4.7\% lower bulk modulus at the core pressure regime with respect to PREM data. The higher bulk modulus estimated in our study might be due to the presence of Ni in the crystal, as Ni is less compressible than Fe\cite{hirao2022equations}. 
\newpage
\subsection{Density estimation}
The density of the synthesized material is calculated by dividing the total atomic mass associated with a unit cell by the unit cell volume. The unit cell volume at core pressure is derived by extrapolating the fitted BM EOS curve up to that pressure regime. The error in volume at core pressure is estimated from the error associated with K$_{0}$ and K$^{'}$. For error estimation in mass, we have used the same percentage of error obtained for the constituent elements during the weighing of our samples.
The estimated density of our sample at core pressure and room temperature (300K) is found to be 13.257 gm/cm$^{3}$ with a maximum error of  $\pm$4.3\%, which is 1.35 \% higher than PREM data. The inner core temperature ranges from 5000 to 7000 K, as estimated by Boehler\cite{boehler1996melting}.To estimate the density of this material at core pressure and temperature conditions, the temperature effect on the density needs to be considered. The effects of temperature were assessed using the Mie-Grüneisen-Debye (MGD) formulation\cite{gruneisen1959state}. Using MGD model actual pressure exerted on the sample can be corrected from the effect of thermal pressure calculated from the equation of state (EOS) of MgO using the following.
\begin{equation}
	P(V,T)=P(V,T_{0})+\Delta P_{th}(V,T)
\end{equation}
where P(V, To) is the pressure component at a reference temperature T$_{0}$, fixed at 300 K, and $\Delta$P$_{th}$ is the thermal pressure produced by the change in temperature from T$_{0}$ at constant volume. 
Thermal pressures  $\Delta$P$_{th}$ is expressed by the variation of internal thermal energies $\Delta$E$_{th}$  between T$_{0}$ and T as\cite{tange2009unified} \begin{equation} \Delta P_{th}(V,T)=P_{th}(V,T)-P_{th}(V,T_{0})=\frac{\gamma (V)}{V}\left [ E_{th}(V,T)-E_{th}(V,T_{0}) \right ]
\end{equation}
where $\gamma$ is the Grüneisen parameter and E$_{th}$ is given by the Debye model as 
\begin{equation}
	E_{th}(V,T)=\frac{9nRT}{\left ( \frac{\Theta_{D} }{T} \right )^{3}}\int_{0}^{\frac{\Theta_{D} }{T}}\frac{t^{3}}{e^{t}-1}dt
\end{equation}
Where R is the universal gas constant =8.314 JK$^{-1}$mol$^{-1}$, n is the no of atoms in the formula unit of the concerned material. Here for MgO, n=2. $\Theta_{D}$ is the Debye temperature. $\Theta_{D}$ is calculated following the expression
\begin{equation}	
	\Theta_{D} =\Theta _{0}exp\left [ \frac{\gamma _{0}-\gamma }{q} \right ]
\end{equation}
Where $\Theta_{0}$ is the volume-independent Debye temperature; $\gamma_{0}$ and$\gamma$ are volume-independent and volume-dependent Grüneisen parameters respectively, q is given by $q=\frac{\partial ln\gamma }{\partial lnV}$.
For MgO at ambient pressure and room temperature $\gamma _{0}$=1.55, $\theta_{0}$ = 773K and  q = 1.65\cite{speziale2001quasi}. For MgO, the volume at ambient pressure and room temperature is V$_{0}$=74.7248(7) \AA$^{3}$. The pressure vs unit cell volume of MgO is fitted using 3$^{rd}$ order Birch Murnaghan EOS and extrapolated to 364 GPa to get V of MgO at core pressure.

The maximum thermal pressure calculated at 5000K and 7000 K are 5.2 GPa and 7.5 GPa, respectively. The pressure values of the data from 43 GPa to 120 GPa are corrected by adding thermal pressure term P$_{th}$ using Eq.(1). Thus P$_{th}$ corrected separate pressure vs unit cell volume data sets are obtained at 5000K and 7000K. At room temperature, the pressure evolution of unit cell volume up to core pressure(364 GPa) is obtained by extrapolating the EOS fitted for region II. The P$_{th}$ corrected pressure vs volume data at 5000K and 7000K are fitted using 3$^{rd}$ order BM EOS by varying V$_{0}$, K$_{0}$ and K$^{'}$. Corresponding fitted EOS for 5000K and 7000K are similarly extrapolated to 364 GPa to get the pressure evolution of the unit cell volume of the sample at corresponding temperatures. 
The pressure evolution of density at 300K, 5000K, and 7000K temperatures are estimated from the separate fitted EOS and are shown in Fig.7. With increasing temperature the density profile approaches the PREM data.
The density of the earth's inner core as per PREM data is 13.088 gm/cm$^{3}$. In our study, the density of (Fe,Ni)$_{7}$(C,Si)$_{3}$ at core pressure and at 5000K is calculated to be 13.258 gm/cm$^{3}$  and at 7000 K it is 13.207 gm/cm$^{3}$, which are only 1.3 \% and 0.9 \% higher than the PREM data, respectively. The density of different possible earth core materials at core pressure are compared with our study at 300K and 7000K, in Fig.8. We can see iron\cite{dubrovinsky2000situ} and iron-nickel alloys\cite{hirao2022equations} have higher densities, whereas studies on Fe$_{7}$C$_{3}$ by Liu et al.\cite{liu2016phase} and Chen et al.\cite{chen2012magneto} and si doped Fe$_{7}$C$_{3}$ by Das et al.\cite{das2017first} show much lower densities than PREM data. The density estimated by Saha et al. for Fe$_{7}$(C,Si)$_{3}$ at 7000K and core pressure is found to be 2\% higher than PREM data. 
The excellent match of density calculated at 364 GPa and 7000K in our study with respect to PREM data shows that Ni can be an important constituent of the Earth's core. Further experimental and theoretical works are necessary to confirm the orthorhombic structure at the core condition.
\newpage

\section{Conclusion}
We have synthesized o-(Fe,Ni)$_{7}$(C,Si)$_{3}$ at high pressure and high temperature using LHDAC. The synthesized sample is characterized using Raman, SAED, and X-ray diffraction measurements. The high-pressure XRD study shows no structural transition in the synthesized material up to around 121 GPa at room temperature. The pressure evolution of unit cell volume shows an anomaly around 79 GPa. The fitted BM EOSs show a slight decrease in the bulk modulus at that pressure region, which might be associated with the magnetic transition present in the material. The decrease in bulk modulus around 79 GPa is found to be less compared to other similar studies, which might be due to the 6.5\% doping of Ni, which shows no magnetic transition at high pressure. The estimated bulk modulus and density at the core pressure region are found to be very close to the PREM data. The presence of Ni, which poses higher stiffness compared to iron might be the reason for the higher bulk modulus of the material compared to other studies. The thermal pressure corrected density at room temperature and 7000K are found to be an excellent match with the PREM data.
\newline
{\bf Acknowledgments}
The authors gratefully acknowledge the Ministry of Earth Sciences, Government of India, for the financial support under grant No. MoES/16/25/10-RDEAS to carry out this high-pressure high-temperature research work. B.Ghosh also gratefully acknowledge the Department of Science and Technology, Government of India for their INSPIRE fellowship grant for pursuing
Ph.D. program. The authors acknowledge DESY (Hamburg, Germany), a member of the Helmholtz Association HGF, for the provision of experimental facilities. Parts of this research were carried out at PETRA-III and they would like to thank beamline scientists for assistance in using the P02.2 beamline. The authors also gratefully acknowledge the financial support from the Department of Science and Technology, Government of India to visit PETRA beamline at DESY, Hamburg, Germany. 

\bibliographystyle{apsrev4-2}
\bibliography{References}
\newpage

\begin{figure}
	\centering
	\includegraphics[width=\columnwidth, height=15cm]{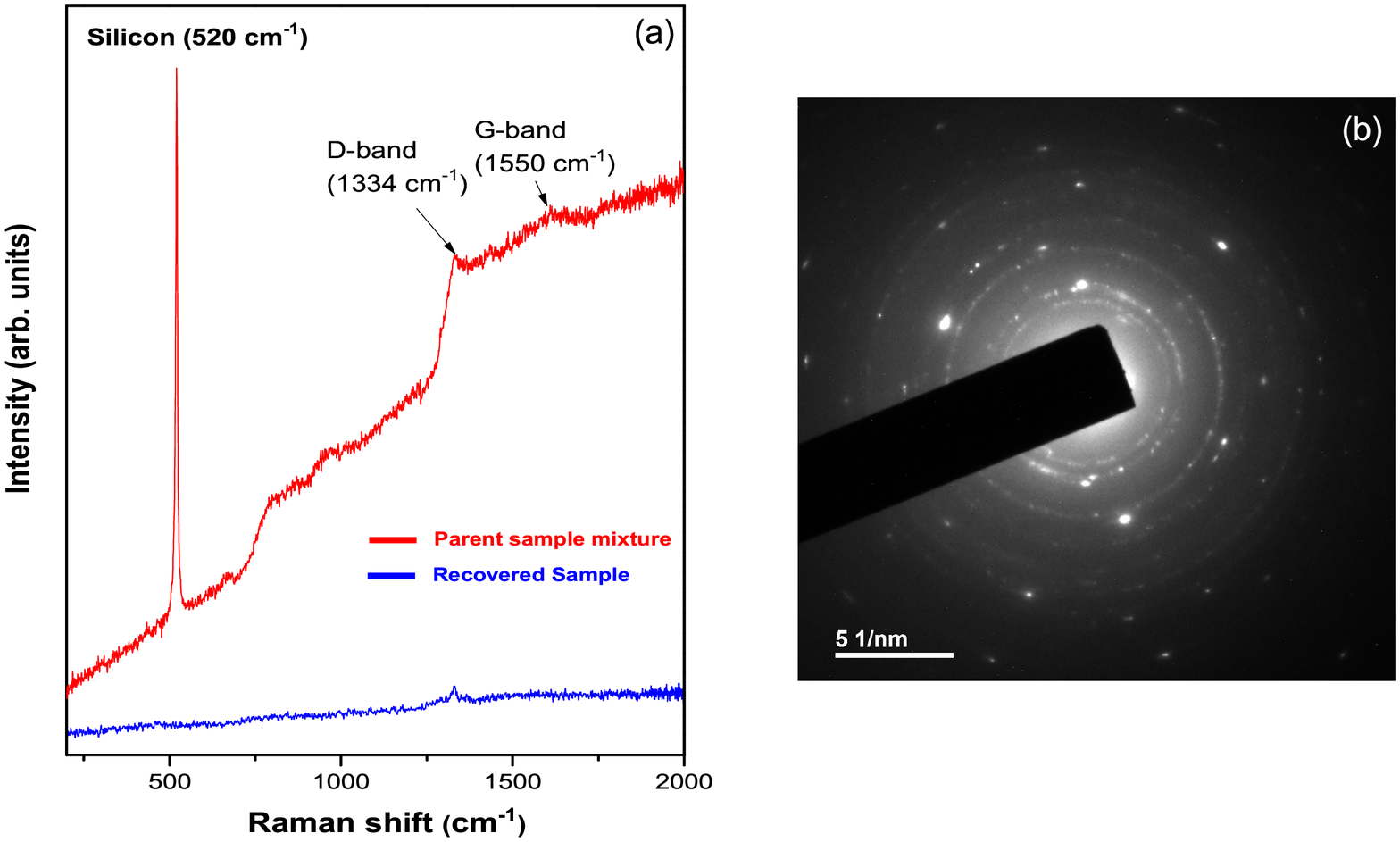}
	\caption{\label{Fig.1}(a) The Raman spectra of the parent mixture sample and recovered synthesized material. (b) Single area electron diffraction pattern of the recovered heated sample. 
	  }
\end{figure}

\begin{figure}
	\centering
	\includegraphics[width=16cm,height=12cm]{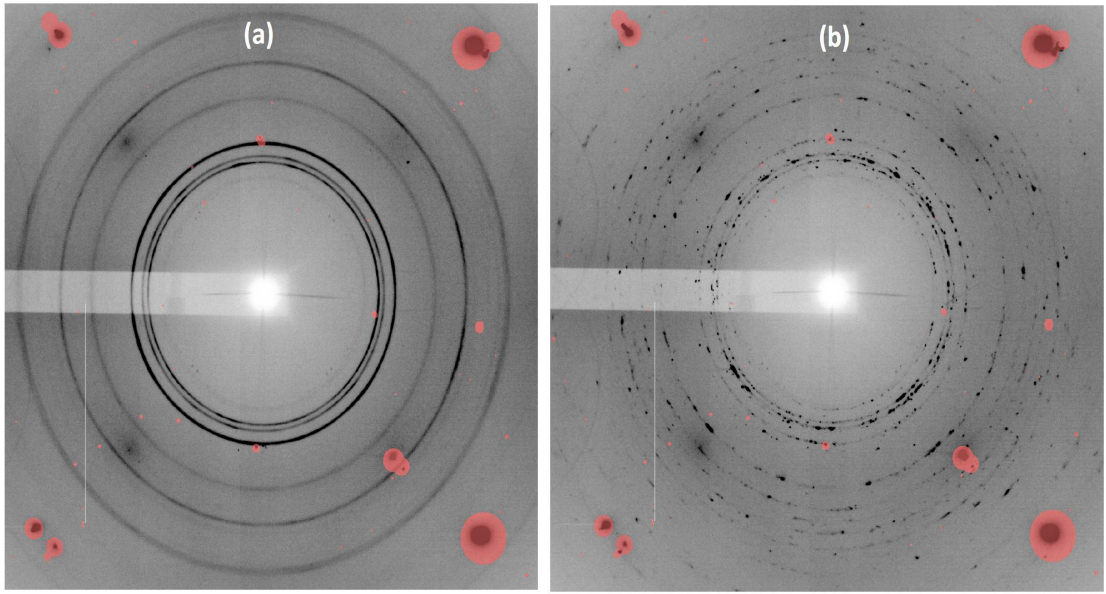}
	\caption{\label{Fig.2} Comparison of 2D XRD images before(a) and after(b) heating. Reflections masked by red correspond to diamond reflection.}
\end{figure} 

\begin{figure}
	\centering
	\includegraphics[width=\columnwidth, height=22cm]{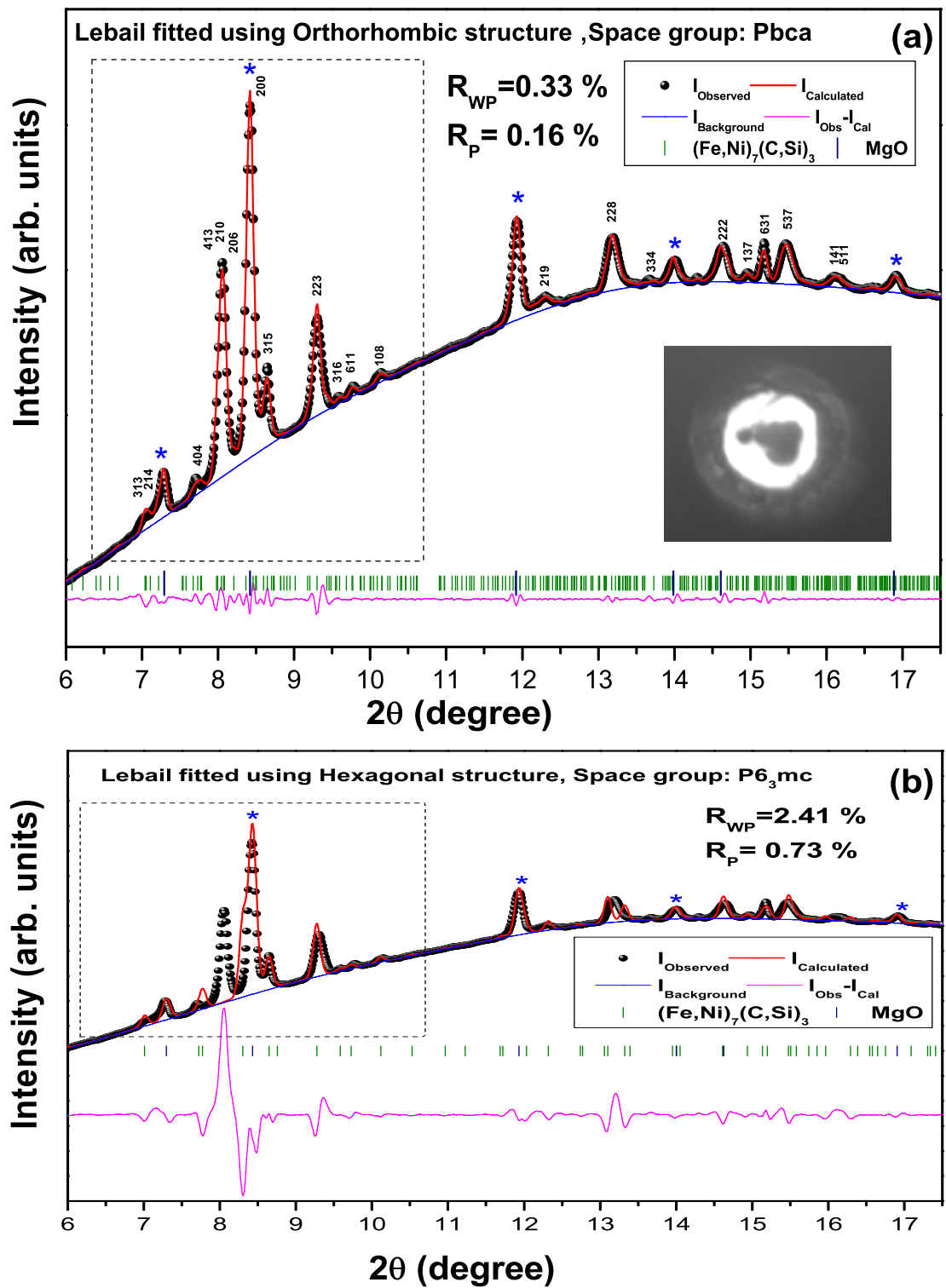}
	\caption{\label{Fig.3} Lebail fitting of the XRD pattern after heating using (a) orthorhombic crystal structure and (b) using hexagonal crystal structure. }
\end{figure}

\begin{figure}
	\centering
	\includegraphics[width=\columnwidth,height=22cm]{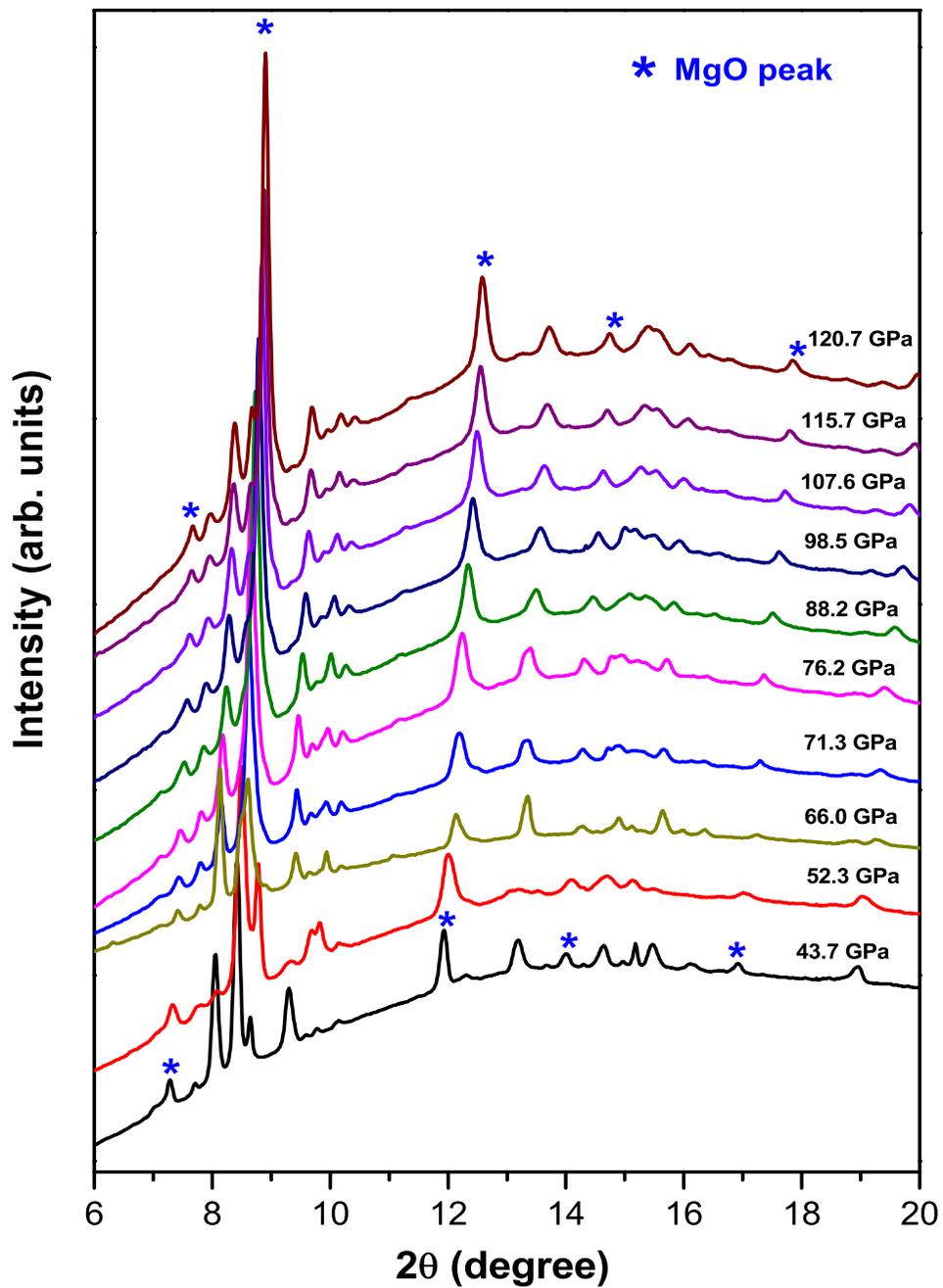}
	\caption{\label{Fig.4} Pressure evolution of XRD patterns after synthesis. }
\end{figure}

\begin{figure}
	\begin{center}
		\includegraphics[width=17cm,height=14cm]{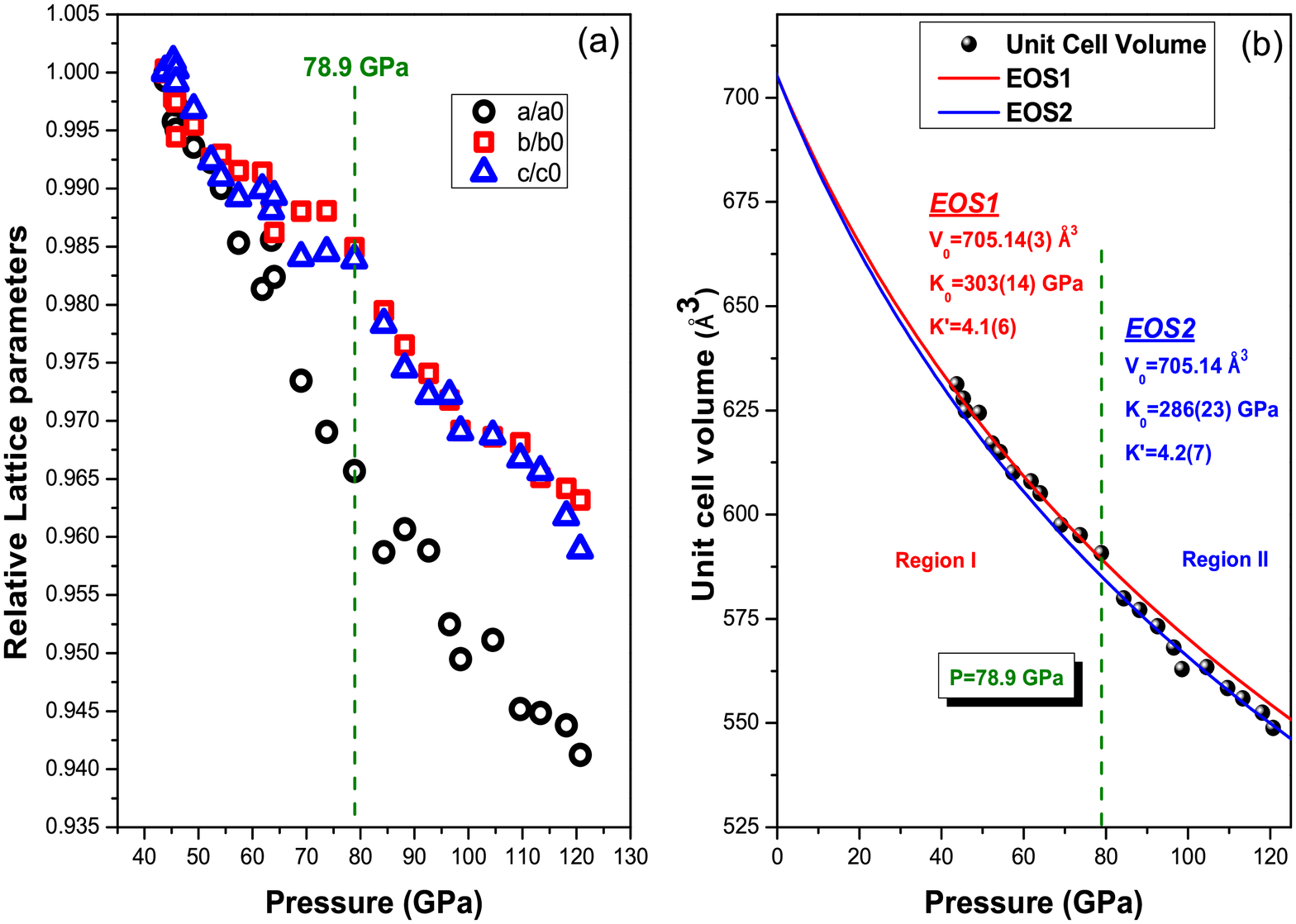}
	\end{center}
	\caption{\label{Fig.5} (a) Pressure evolution of relative lattice parameters. (b) Pressure evolution of unit cell volume. Pressure vs unit cell volume data is fitted using two separate 3$^{rd}$ order Birch Murnaghan EOSs in two pressure regions. }
\end{figure}

\begin{figure}
	\centering
	\includegraphics[width=17cm, height=14cm]{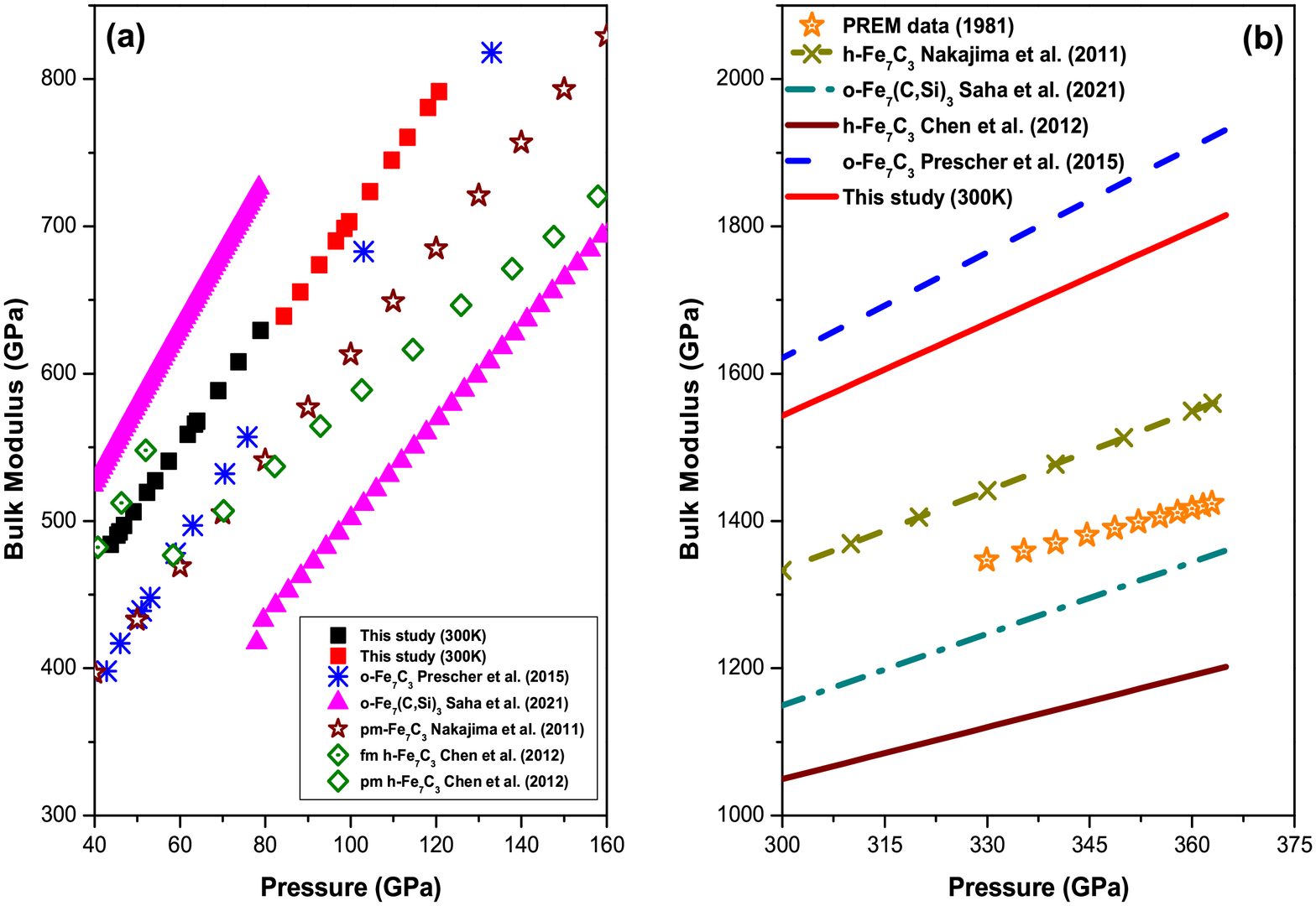}
	\caption{\label{Fig.6} (a) Comparison of pressure variation of bulk modulus with other studies. (b) Comparison of pressure variation of extrapolated bulk modulus with other studies and PREM data at core pressure region.}
\end{figure}

\begin{figure}
	\centering
	\includegraphics[width=\columnwidth,height=20cm]{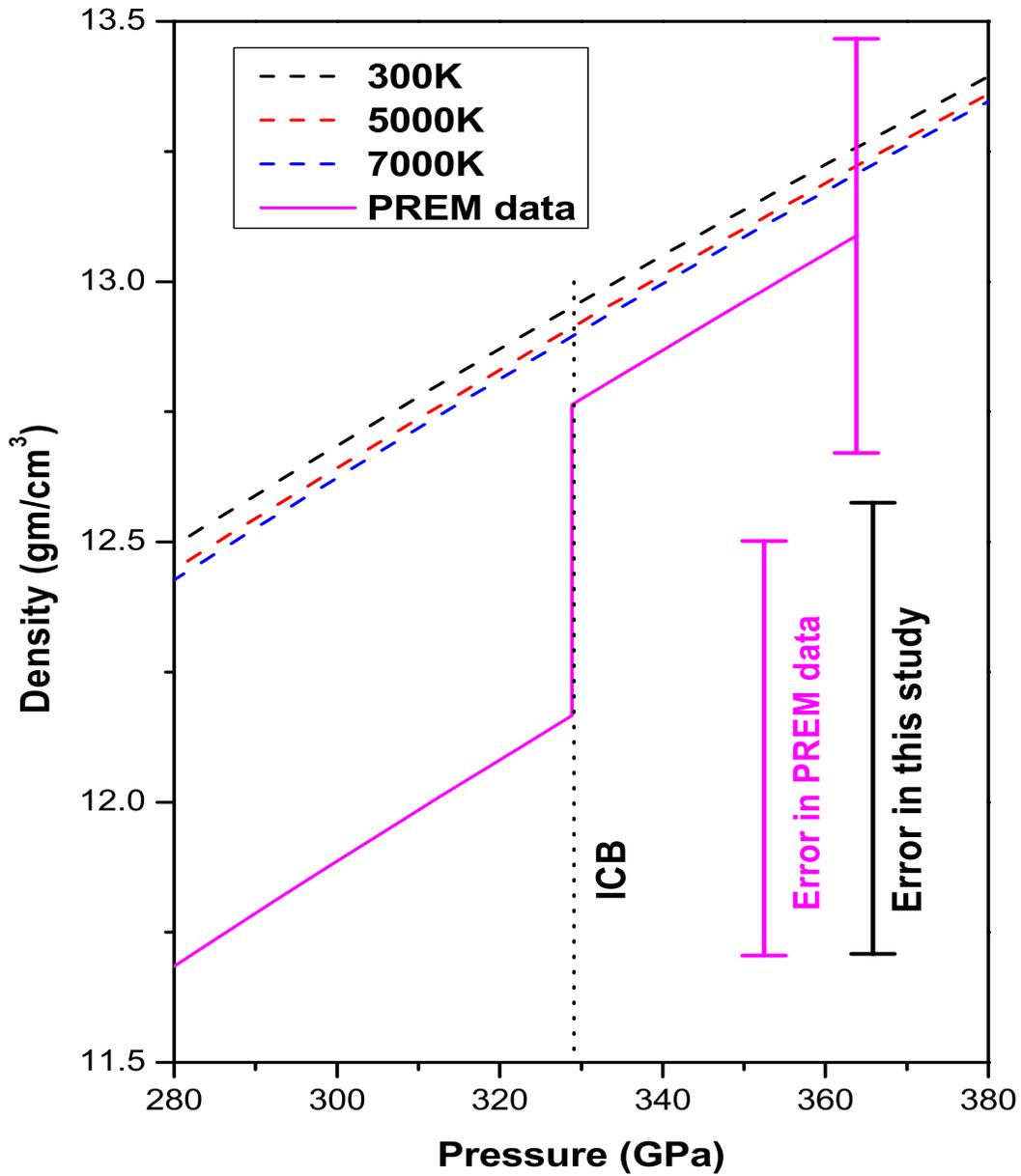}
	\caption{\label{Fig.7} Pressure evolution of density at room temperature as well as at 5000K and 7000K. Dotted vertical line signifies the inner core boundary.}
\end{figure}

\begin{figure}
	\centering
	\includegraphics[width=\columnwidth , height=22cm]{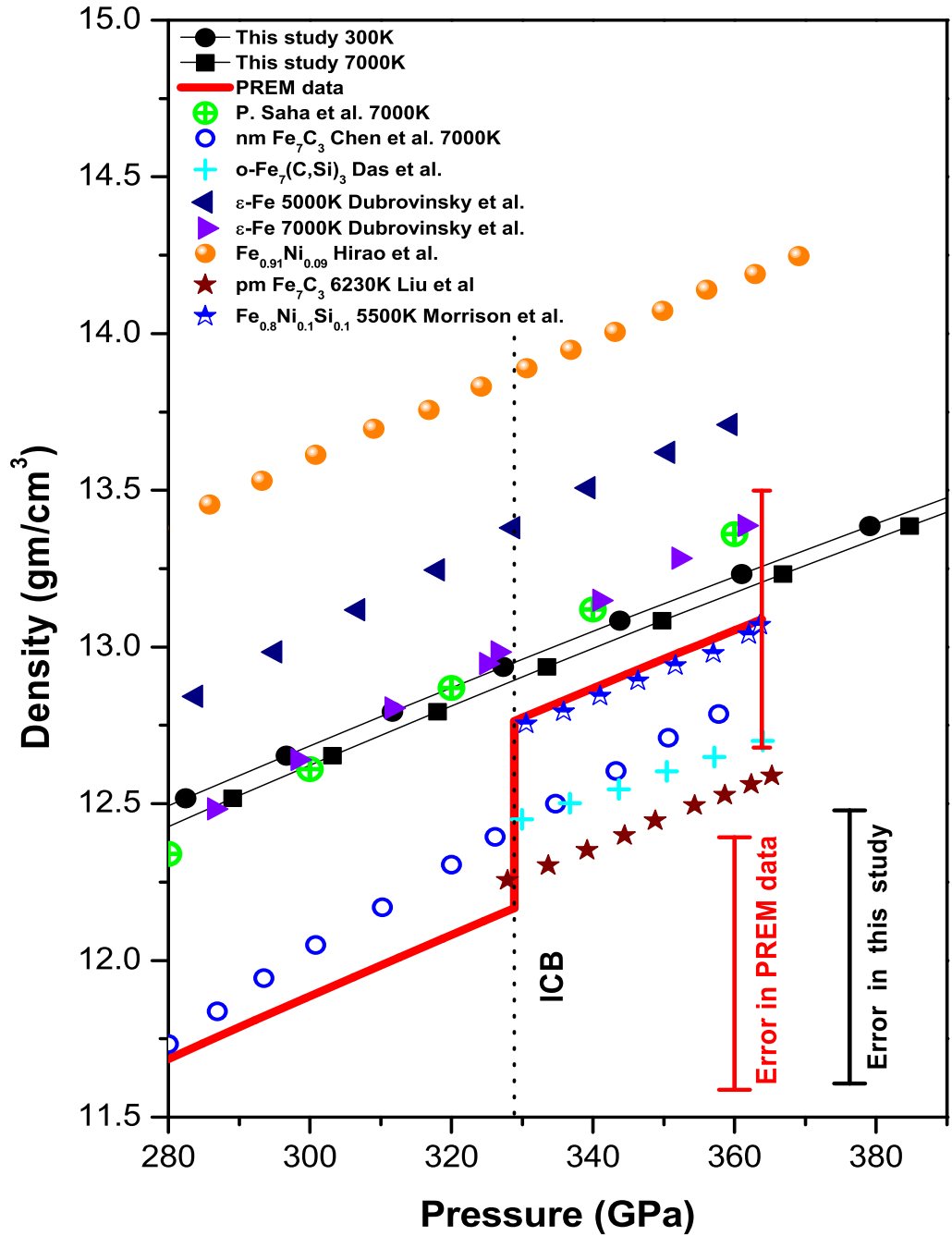}
	\caption{\label{Fig.8} Comparison of estimated density variation at core pressure region between different studies. }
\end{figure}

\end{document}